# Self-Dual Yang-Mills and Vector-Spinor Fields, Nilpotent Fermionic Symmetry, and Supersymmetric Integrable Systems

Hitoshi Nishino[1]) and Subhash Rajpoot[2])

*Department of Physics & Astronomy*
*California State University*
*1250 Bellflower Boulevard*
*Long Beach, CA 90840*

## Abstract

We present a system of a self-dual Yang-Mills field and a self-dual vector-spinor field with nilpotent fermionic symmetry (but not supersymmetry) in $2+2$ dimensions, that generates supersymmetric integrable systems in lower dimensions. Our field content is $(A_\mu{}^I, \psi_\mu{}^I, \chi^{IJ})$, where $I$ and $J$ are the adjoint indices of arbitrary gauge group. The $\chi^{IJ}$ is a Stueckelberg field for consistency. The system has local nilpotent fermionic symmetry with the algebra $\{N_\alpha{}^I, N_\beta{}^J\} = 0$. This system generates supersymmetric Kadomtsev-Petviashvili equations in $D = 2+1$, and supersymmetric Korteweg-de Vries equations in $D = 1+1$ after appropriate dimensional reductions. We also show that a similar self-dual system in seven dimensions generates self-dual system in four dimensions. Based on our results we conjecture that lower-dimensional supersymmetric integral models can be generated by non-supersymmetric self-dual systems in higher dimensions only with nilpotent fermionic symmetries.

[1]) E-Mail: hnishino@csulb.edu
[2]) E-Mail: rajpoot@csulb.edu

## 1. Introduction

The mathematical conjecture that self-dual Yang-Mills theory in $D = 2+2$ space-time dimensions is likely to be the master theory for all integrable models in lower dimensions [1] has received much attention in physics community. One of the reasons is that Ooguri and Vafa [2] showed that the consistent backgrounds for $N = 2$ string theory should be self-dual gravity field for closed $N = 2$ strings, self-dual Yang-Mills field for open strings, and self-dual Yang-Mills plus gravity in the case of $N = 2$ heterotic strings in $D \leq 4$. Also, topological strings are known to unify non-critical (super)strings, integrable models, and matrix models [3].

These developments elucidate the importance of self-dual supersymmetric Yang-Mills models in $D = 2+2$ [4][5]. The common notion is that the most fundamental $N = 1$ self-dual supersymmetric Yang-Mills multiplet should contain the spins $(1, 1/2)$. However, the supermultiplet $(1, 1/2)$ for self-dual supersymmetric Yang-Mills may be not unique, because of an alternative spin content $(3/2, 1)$ with a vector-spinor $\psi_\mu$.[3)] However, an interacting $(3/2, 1)$ multiplet seems to imply local supersymmetry, because the index $\mu$ on $\psi_\mu$ requires the transformation $\delta_Q \psi_\mu = \partial_\mu \epsilon + \cdots$ for consistent gauge interactions. Then the closure of two supersymmetries leads to the space-time dependent parameter $\xi^\mu = (\overline{\epsilon}_1 \gamma^\mu \epsilon_2)$ for local translational symmetry necessitating a graviton, and thereby supergravity [6]. So there seems to be no consistent way of introducing a vector-spinor as a super-partner field for self-dual Yang-Mills field without supergravity.

One way to avoid this problem is as follows. We do not have to maintain 'supersymmetry' in $D = 2+2$. For example, as in [7] only nilpotent fermionic symmetry may be realized in $D = 2+2$, whereas supersymmetries in $D \leq 3$ may emerge as hidden symmetries. In the present paper, we present such a system with the same field content $(A_\mu{}^I, \psi_\mu{}^I, \chi^{IJ})$ as in [7]. Local nilpotent fermionic symmetry is needed in [7] for consistency of the total system. We present a self-dual Yang-Mills field and a self-dual vector-spinor with nilpotent fermionic symmetry, generating supersymmetric integrable systems in $D \leq 3$ after dimensional reductions. We also propose similar theories in $D \geq 5$, based on the 'generalized' self-

---
[3)] In supergravity [6], $\psi_\mu$ is called 'gravitino' as the super-partner of the graviton $g_{\mu\nu}$. In this paper, we use the phrase 'vecor-spinor', avoiding the word 'gravitino' which is the super-partner of the graviton.

duality.

We stress in this paper the existence of 'hidden' supersymmetries that is not manifest in the original 4D. We use the terminology 'hidden', because supersymmetries in dimensions in $D = 2+1$ or $D = 1+1$ arising after dimensional reductions are not manifest in the original 4D. This situation is in a sense very similar to the hidden $E_{7(+7)}/SU(8)$ symmetry in $N = 8$ supergravity in 4D [8]. Even though the final $E_{7(+7)}/SU(8)$ symmetry in 4D is supposed to be a part of the original $N = 1$ supergravity system in 11D, such symmetry is not manifest, at least in a Lorentz-covariant manner in the original 11D.

This paper is organized as follows. In the next section, we give the foundation of our system based on [7]. In section 3, we give the special case of $D = 2+2$, and give the explicit forms of self-duality conditions. We also prepare for dimensional reductions into lower dimensions. In section 4, we perform the dimensional reduction into $D = 2+1$, and show that $N = 1$ supersymmetric supersymmetric Kadomtsev-Petviashvili equations are generated. Similarly, in section 5, we perform a dimensional reduction into $D = 1+1$, and show that $N = 1$ supersymmetric Korteweg-de Vries equations are generated. In section 6, we give a similar system in 7D with generalized self-duality. This system can be regarded as a more fundamental system than 4D system, because the former generates the latter by a simple dimensional reduction.

## 2. Foundation of System

We start with our algebra in the system [7]:[4)]

$$\left\{N_{\underline{\alpha}}{}^I, N_{\underline{\beta}}{}^J\right\} = 0 \quad, \quad \left[T^I, N_{\underline{\alpha}}{}^J\right] = +f^{IJK} N_{\underline{\alpha}}{}^K \quad, \quad \left[T^I, T^J\right] = +f^{IJK} T^K \quad, \quad (2.1)$$

where $I, J, \cdots = 1, 2, \cdots, \dim G$ are the adjoint index for a Yang-Mills gauge group $G$. The $N_{\underline{\alpha}}{}^I$ are the nilpotent fermionic generators, while $T^I$ are the usual anti-hermitian generators for the group $G$. We use $\underline{\alpha}, \underline{\beta}, \cdots = \underline{1}, \underline{2}, \underline{3}, \underline{4}$ as the spinorial index for a Majorana spinors in

---

[4)] We use the symbol $N_{\underline{\alpha}}{}^I$ for the nilpotent fermionic generator, lest readers should confuse it with the generator $Q_{\underline{\alpha}}$ of supersymmetry.

$D = 2+2$ [4][5][5)] As in [7], the corresponding field strengths are [7]

$$F_{\mu\nu}{}^I \equiv +\partial_\mu A_\nu{}^I - \partial_\nu A_\mu{}^I + f^{IJK} A_\mu{}^J A_\nu{}^K \ , \tag{2.2a}$$

$$\mathcal{R}_{\mu\nu}{}^I \equiv +D_\mu \psi_\nu{}^I - D_\nu \psi_\mu{}^I + \chi^{IJ} F_{\mu\nu}{}^J \ , \tag{2.2b}$$

$$\mathcal{D}_\mu \chi^{IJ} \equiv +\partial_\mu \chi^{IJ} + 2 f^{[I|KL} A_\mu{}^K \chi^{L|J]} + f^{IJK} \psi_\mu{}^K \equiv +D_\mu \chi^{IJ} + f^{IJK} \psi_\mu{}^K \ , \tag{2.2c}$$

where $D_\mu$ is the gauge-covariant derivative. The peculiar Chern-Simons terms in (2.2b) and (2.2c) are needed for the invariance of these field strengths [7]. The $\psi_\mu{}^I$ and $\chi^{IJ}$ are 2-component Majorana-Weyl spinors in $D = 2+2$ composed of one-component spinors:

$$\psi_\mu{}^I \equiv \begin{pmatrix} \lambda_\mu{}^I \\ \lambda_\mu{}^{I*} \end{pmatrix} \ , \quad \chi^{IJ} \equiv \begin{pmatrix} \omega^{IJ} \\ \omega^{IJ*} \end{pmatrix} \ , \tag{2.3}$$

where $*$ implies a complex conjugate [4][5]. The Bianchi identities for (2.2) are $D_{[\mu} F_{\nu\rho]}{}^I \equiv 0$ and [7]

$$D_{[\mu} \mathcal{R}_{\nu\rho]}{}^I \equiv +F_{[\mu\nu}{}^J \mathcal{D}_{\rho]} \chi^{IJ} \ , \quad D_{[\mu} \mathcal{D}_{\nu]} \chi^{IJ} \equiv +\tfrac{1}{2} f^{IJK} \mathcal{R}_{\mu\nu}{}^K - \tfrac{3}{2} f^{[IJ|K} F_{\mu\nu}{}^L \chi^{K|L]} \ . \tag{2.4}$$

Our nilpotent fermionic transformation $\delta_N$ is

$$\delta_N \psi_\mu{}^I = +D_\mu \zeta^I \ , \quad \delta_N A_\mu{}^I = 0 \ , \quad \delta_N \chi^{IJ} = -f^{IJK} \zeta^K \ . \tag{2.5a}$$

$$\delta_N F_{\mu\nu}{}^I = 0 \ , \quad \delta_N \mathcal{R}_{\mu\nu}{}^I = 0 \ , \quad \delta_N (\mathcal{D}_\mu \chi^{IJ}) = 0 \ , \tag{2.5b}$$

where $\zeta^{\underline{\alpha} I}$ is the parameter for the nilpotent fermionic symmetry $N_{\underline{\alpha}}$. Our fields are also transforming appropriately under the gauge transformation $\delta_T$:

$$\delta_T (A_\mu{}^I, \psi_\mu{}^I, \chi^{IJ}) = (+D_\mu \Lambda^I, -f^{IJK} \Lambda^J \psi_\mu{}^K, -2 f^{[I|K} \Lambda^K \chi^{L|J]}) \ , \tag{2.6a}$$

$$\delta_T (F_{\mu\nu}{}^I, \mathcal{R}_{\mu\nu}{}^I, \mathcal{D}_\mu \chi^{IJ}) = (-f^{IJK} \Lambda^J F_{\mu\nu}{}^K, -f^{IJK} \Lambda^J \mathcal{R}_{\mu\nu}{}^K, -2 f^{[I|KL} \Lambda^K \mathcal{D}_\mu \chi^{L|J]}) \ , \tag{2.6b}$$

showing the consistency of the system. For example, we can not skip the last terms in (2.2b) and (2.2c), because they will lead to non-invariance of the field strengths in (2.6) [7].

The closure of gauge algebra is also confirmed as

$$[\delta_N(\zeta_1), \delta_N(\zeta_2)] = 0 \ , \tag{2.7a}$$

$$[\delta_N(\zeta), \delta_T(\Lambda)] = \delta_N(\zeta_3) \ , \quad \zeta_3^I \equiv -f^{IJK} \Lambda^J \zeta^K \ , \tag{2.7b}$$

$$[\delta_T(\Lambda_1), \delta_T(\Lambda_2)] = \delta_T(\Lambda_3) \ , \quad \Lambda_3^I \equiv +f^{IJK} \Lambda_1^J \Lambda_2^K \ . \tag{2.7c}$$

---

[5)] Actually, the formulae in (2.2) through (2.7) except for (2.3) are valid in arbitrary space-time dimensions, not limited to $D = 2+2$, as has been also mentioned in [7].

Note that the properties of our system $(A_\mu{}^I, \psi_\mu{}^I, \chi^{IJ})$ established so far except for (2.3) are valid also in arbitrary space-time dimensions $D$, as has been also explicitly stated in [7]. We can also generalize the space-time signatures to arbitrary ones.

## 3. Space-Time Dimensions $D = 2+2$ and Hidden Supersymmetry

We now limit our space-time dimensions to be $D = 2 + 2$. We impose self-duality conditions on the $F$ and $\mathcal{R}$-field strengths as[6]

$$F_{\mu\nu}{}^I \stackrel{*}{=} +\tfrac{1}{2}\epsilon_{\mu\nu}{}^{\rho\sigma}F_{\rho\sigma}{}^I , \tag{3.1a}$$

$$\mathcal{R}_{\mu\nu}{}^I \stackrel{*}{=} +\tfrac{1}{2}\epsilon_{\mu\nu}{}^{\rho\sigma}\mathcal{R}_{\rho\sigma}{}^I . \tag{3.1b}$$

Needless to say, these self-dualities are also consistent with our nilpotent fermionic symmetry (2.5), because each field strength is invariant under $\delta_N$.

For generating supersymmeric integrable systems in later sections, we use the special metric [9][10],

$$ds^2 = 2dz\,dx + 2dy\,dt . \tag{3.2}$$

In terms of these coordinates, our self-duality (3.1) is

$$F_{xt}{}^I \stackrel{*}{=} 0 , \quad F_{yz}{}^I \stackrel{*}{=} 0 , \quad F_{zx}{}^I \stackrel{*}{=} +F_{ty}{}^I , \tag{3.3a}$$

$$\mathcal{R}_{xt}{}^I \stackrel{*}{=} 0 , \quad \mathcal{R}_{yz}{}^I \stackrel{*}{=} 0 , \quad \mathcal{R}_{zx}{}^I \stackrel{*}{=} +\mathcal{R}_{ty}{}^I . \tag{3.3b}$$

We use also the symbols for fields

$$A_t \equiv H , \quad A_x \equiv Q , \quad A_y \equiv P , \quad A_z \equiv B , \tag{3.4a}$$

$$\lambda_t \equiv \tau , \quad \lambda_x \equiv \xi , \quad \lambda_y \equiv \eta , \quad \lambda_z \equiv \zeta . \tag{3.4b}$$

The spinor field $\lambda$ is upper one-component spinor in (2.3). Each field carries generators $T^I$ implicitly, e.g. $A_t \equiv A_t{}^I T^I$. Our self-duality (3.1) is equivalent to

$$\partial_x H - \partial_t Q + [Q, H] \stackrel{*}{=} 0 , \tag{3.5a}$$

---

[6] We use the symbol $\stackrel{*}{=}$ for an equality that holds upon self-duality conditions or certain ansätze for dimensional reductions.

$$\partial_y B - \partial_z P + [P,\, B] \stackrel{*}{=} 0 ~, \tag{3.5b}$$

$$\partial_z Q - \partial_x B - \partial_t P + \partial_y H + [B,\, Q] - [H,\, P] \stackrel{*}{=} 0 ~, \tag{3.5c}$$

$$\partial_x \tau - \partial_t \xi + [Q,\, \tau] - [H,\, \xi] \stackrel{*}{=} 0 ~, \tag{3.5d}$$

$$\partial_y \zeta - \partial_z \eta + [P,\, \zeta] - [B,\, \eta] \stackrel{*}{=} 0 ~, \tag{3.5e}$$

$$\partial_z \xi - \partial_x \zeta - \partial_t \eta + \partial_y \tau + [B,\, \xi] - [Q,\, \zeta] - [H,\, \eta] + [P,\, \tau] \stackrel{*}{=} 0 ~. \tag{3.5f}$$

We can show that the system (3.5) has hidden supersymmetry. This hidden supersymmetry should not be confused with our original nilpotent fermionic symmetry $N_{\underline\alpha}$. We use the word 'hidden', because the supersymmetry we are going to discuss is not manifest realized in Lorentz-covariant way in the original $D = 2+2$. Such hidden supersymmetry is realized after breaking the original Lorentz symmetry in $D = 2+2$, for the purpose of dimensional reductions.

The explicit form of hidden supersymmetry is dictated by

$$\delta_\alpha H = (\alpha\,\tau) ~, \quad \delta_\alpha Q = (\alpha\,\xi) ~, \quad \delta_\alpha P = (\alpha\,\eta) ~, \quad \delta_\alpha B = (\alpha\,\zeta) ~,$$

$$\delta_\alpha \tau = \alpha\,\partial_x H ~, \quad \delta_\alpha \xi = \alpha\,\partial_x Q ~, \quad \delta_\alpha \eta = \alpha\,\partial_x P ~, \quad \delta_\alpha \zeta = \alpha\,\partial_x B ~, \tag{3.6}$$

where $\alpha$ is one-component spinor. The closure of supersymmetry is

$$[\delta_{\alpha_1},\, \delta_{\alpha_2}] = 2(\alpha_1 \alpha_2)\, \partial_x ~. \tag{3.7}$$

It is straightforward to confirm that the equations in (3.5) are consistent under supersymmetry (3.6). It is clear that hidden the supersymmetry (3.6) breaks the original Lorentz symmetry in $D = 2+2$. Therefore, the meaning of 'hidden' supersymmetry is also evident, because to realize such supersymmetry, the original Lorentz symmetry in $D = 2+2$, such as limiting the parameter of supersymmetry to be one-component spinor, and the direction of translation to be only $\partial_x$.

Eq. (3.5) has another kind of hidden supersymmetry, iff all the fields are Abelian:

$$\delta_\beta H = (\beta\,\partial_x \tau) ~, \quad \delta_\beta Q = (\beta\,\partial_x \xi) ~, \quad \delta_\beta P = (\beta\,\partial_x \eta) ~, \quad \delta_\beta B = (\beta\,\partial_x \zeta) ~,$$

$$\delta_\beta \tau = \beta H ~, \quad \delta_\beta \xi = \beta Q ~, \quad \delta_\beta \eta = \beta P ~, \quad \delta_\beta \zeta = \beta B ~, \tag{3.8}$$

with the one-component spinor $\beta$. The closure of supersymmetry is

$$[\delta_{\beta_1},\, \delta_{\beta_2}] = 2(\beta_1\beta_2)\,\partial_x \quad. \tag{3.9}$$

Out of two supersymmetries (3.6) and (3.8), which one is realized depends on field representations, as will be shown shortly.

## 4. Dimensional Reductions into D=2+1 and Supersymmetric Kadomtsev-Petviashvili Equations

As an explicit application, we perform a dimensional reduction into $D = 2+1$ with the coordinates $(t, x, y)$, and show that $N = 1$ supersymmetric Kadomtsev-Petviashvili equations [11]

$$\tfrac{3}{4}\,\partial_y^2 u + \partial_x\left[\,\partial_t u + \tfrac{1}{4}\,\partial_x^3 u + 3u\,\partial_x u - \tfrac{3}{2}\,\phi\,\partial_x^2\phi\,\right] \doteq 0 \quad, \tag{4.1a}$$

$$\tfrac{3}{4}\,\partial_y^2 \phi + \partial_x\left[\,\partial_t \phi + \tfrac{1}{4}\,\partial_x^3\phi + \tfrac{3}{2}\,\partial_x(u\phi)\,\right] \doteq 0 \quad. \tag{4.1b}$$

are generated. Here $u$ is a real scalar, and $\phi$ is a one-component fermion. Eq. (4.1) is re-expressed as[7]

$$\tfrac{3}{4}\,\partial_y^2\Psi \doteq -\partial_x\left[\,\partial_t\Psi + \tfrac{1}{4}\,\partial_x^3\Psi + \tfrac{3}{2}\,\partial_x(\Psi D\Psi)\,\right] \quad, \tag{4.2}$$

in terms of a fermionic $N = 1$ superfield $\Psi(t, x, y, \theta)$ [11]:

$$\Psi(t,x,y,\theta) \equiv \phi(t,x,y) + \theta\,u(t,x,y)\quad, \qquad D \equiv \partial_\theta + \theta\partial_x \quad, \qquad D^2 = \partial_x \quad. \tag{4.3}$$

This dimensional reduction into $D = 2+1$ is performed by the ansätze

$$\partial_z \stackrel{*}{=} 0 \quad, \qquad B \stackrel{*}{=} 0 \quad, \qquad \zeta \stackrel{*}{=} 0 \quad, \tag{4.4}$$

and

$$H \equiv +\tfrac{3}{4}\,\partial_t\,\partial_y u \quad,\qquad Q \equiv +\tfrac{3}{4}\,\partial_x\,\partial_y u \quad,\qquad P \equiv -\partial_x\left[\,\partial_t u + \tfrac{1}{4}\,\partial_x^3 u + 3u\,\partial_x u - \tfrac{3}{2}\,\phi\,\partial_x^2\phi\,\right] \quad,$$

$$\tau \equiv +\tfrac{3}{4}\,\partial_t\partial_y\phi \quad,\qquad \xi \equiv +\tfrac{3}{4}\,\partial_x\partial_y\phi \quad,\qquad \eta \equiv -\partial_x\left[\,\partial_t\phi + \tfrac{1}{4}\,\partial_x^3\phi + \tfrac{3}{2}\,\partial_x(u\phi)\,\right] \quad. \tag{4.5}$$

---

[7] We use the symbol $\doteq$ for a field equation, for an equality valid by the use of field equation(s), or for an ansatz for dimensional reduction as in section 6.

Each field is Abelian without any generator. All equations in (3.5) are satisfied by (4.5) with (4.4), except (3.5c) and (3.5f), which in turn generate (4.1) with an overall time-derivative $\partial_t$. The integral constant integrating $\partial_t$ is excluded by the boundary condition $\lim_{|x|\to\infty} u(t,x,y) = \lim_{|x|\to\infty} \phi(t,x,y) = 0$.

Our hidden supersymmetry is the dimensionally-reduced version of supersymmetry (3.8) under (4.4):

$$\delta_\beta H = (\beta\,\partial_x \tau) \ , \quad \delta_\beta Q = (\beta\,\partial_x \xi) \ , \quad \delta_\beta P = (\beta\,\partial_x \eta) \ , \tag{4.6a}$$

$$\delta_\beta \tau = \beta H \ , \quad \delta_\beta \xi = \beta Q \ , \quad \delta_\beta \eta = \beta P \ , \tag{4.6b}$$

when $u$ and $\phi$ are transforming as $\delta_\beta \phi = \beta D\Psi\big| = \beta u$, $\delta_\beta u = \beta D(D\Psi)\big| = (\beta\partial_x\phi)$.

## 5. Dimensional Reduction into $D=1+1$ and Supersymmetric Korteweg-de Vries Equations

We next perform a dimensional reduction into $D = 1+1$, and show supersymmetry. We require $\partial_y \stackrel{*}{=} \partial_z \stackrel{*}{=} 0$, and choose $Q \equiv A_x$ and $H \equiv A_t$ to be zero [9][10]. The corresponding components of $\psi_\mu{}^I$ are also put to zero:

$$\partial_y \stackrel{*}{=} \partial_z \stackrel{*}{=} 0 \ , \quad Q \stackrel{*}{=} 0 \ , \quad H \stackrel{*}{=} 0 \ , \quad \xi \equiv \psi_x \stackrel{*}{=} 0 \ , \quad \tau \equiv \psi_t \stackrel{*}{=} 0 \ . \tag{5.1}$$

The self-duality conditions (3.5) under (5.1) are equivalent to the four equations

$$[P,B] \stackrel{*}{=} 0 \ , \quad \dot P \stackrel{*}{=} -B' \ , \quad \dot\eta \stackrel{*}{=} -\zeta' \ , \quad [P,\zeta] \stackrel{*}{=} [B,\eta] \ , \tag{5.2}$$

where a dot (or prime) stands for $\partial_t$ (or $\partial_x$). These conditions agree with those arising from $N=1$ self-dual supersymmetric Yang-Mills in $D=2+2$ [12] (Cf. Eqs. (2.9), (2.10), (2.14) and (2.15) in [12].) This is already evidence that (5.2) has hidden supersymmetry in $D=1+1$. In fact, the system (3.5) has supersymmetry under the dimensional reduction (5.1):

$$\delta P = +(\beta\eta) \ , \quad \delta B = +(\beta\zeta) \ , \quad \delta\eta = +\widetilde\beta P' + \widetilde\alpha \dot P \ , \quad \delta\zeta = +\widetilde\alpha \dot B + \widetilde\beta B' \ . \tag{5.3}$$

The closure of supersymmetry is $[\delta_1, \delta_2] = (\beta_2\widetilde\beta_1 \partial_x + \beta_2\widetilde\alpha_1 \partial_t) - (1 \leftrightarrow 2)$. Note that supersymmetry was not realized in the original space-time $D=2+2$, and therefore supersymmetry (5.3) is unexpectedly larger symmetry compared with the original $D = 2+2$. In other words, we had only $N = 0$ supersymmetry in $D = 2+2$, but after the dimensional reduction, we obtained $N = 1$ supersymmetry as the enlargement of symmetries. This is a new phenomenon occurring in our peculiar system originally in $D = 2+2$ only with nilpotent fermionic symmetry but not supersymmetry. According to common wisdom, supersymmetries are supposed to be broken or at most preserved in dimensional reduction, while our system showed that supersymmetries $N > 0$ arise out of non-supersymmetry $N = 0$ in higher dimensional parental theory. Notice also that the fermionic fields in (5.3) originate from the vector-spinor in the parental theory in $D = 2+2$.

As an explicit example, we consider the $N = 1$ supersymmetric Korteweg-de Vries equations in $D = 1+1$ [13]:[8]

$$\dot{u} \doteq -u''' + 6uu' - 3\phi\phi'' = -(u'' - 3u^2 + 3\phi\phi')' \equiv -f'(x,t) \ , \quad (5.4a)$$

$$\dot{\phi} \doteq -\phi''' + 3u'\phi + 3u\phi' = -(\phi'' - 3u\phi)' \equiv -g'(x,t) \ , \quad (5.4b)$$

where $u$ is a real scalar, and $\phi$ is a one-component spinor. This is equivalent to [13]

$$\dot{\Psi} \doteq -\Psi''' + 3(\Psi D\Psi)' = -(+\Psi'' - 3\Psi D\Psi)' \ , \quad (5.5a)$$

$$\Psi(x,t,\theta) \equiv \phi(x,t) + \theta\, u(x,t) \ , \quad D \equiv \partial_\theta + \theta\partial_x \ , \quad D^2 = \partial_x \ . \quad (5.5b)$$

Eq. (3.5) generates supersymmetric Korteweg-de Vries equations (5.5), under the Abelian-case ansatz

$$\eta \equiv \Psi\,| = \phi \ , \quad P \equiv D\Psi\,| = u \ , \quad (5.6a)$$

$$\zeta \equiv (+\Psi'' - 3\Psi D\Psi)\big| \ , \quad B \equiv [\,D(+\Psi'' - 3\Psi D\Psi)\,]\big| \ . \quad (5.6b)$$

The supersymmetry transformation $\delta_\beta$ in (3.8) is now restricted under the dimensional reduction condition (5.1) as

$$\delta_\beta\phi = \beta D\Psi\big| = \beta u \ , \quad \delta_\beta u = \beta D(D\Psi)\big| = (\beta\partial_x\phi) \ , \quad (5.7a)$$

$$\delta_\beta P = (\beta\,\partial_x\eta) \ , \quad \delta_\beta B = (\beta\,\partial_x\zeta) \ , \quad \delta_\beta\eta = \beta P \ , \quad \delta_\beta\zeta = \beta B \ . \quad (5.7b)$$

---

[8] These equations are called supersymmetric Korteweg-de Vries-3 equation in [13].

## 6. Example of Similar System in 7D

As we have promised, we next give an explicit analog in 7D. In 7D, there are generalized self-duality conditions, based on the so-called octonionic structure constant [14] and reduced $G_2$ holonomy [15][16]. In Euclidian 7D, the reduced holonomy is $G_2$ as the subgroup of the maximal holonomy $SO(7)$ [17]. The explicit form of self-duality condition in 7D on a Yang-Mills field is

$$F_{\mu\nu}{}^I \stackrel{*}{=} + \tfrac{1}{2} \phi_{\mu\nu}{}^{\rho\sigma} F_{\rho\sigma}{}^I ~,\qquad (6.1)$$

where $\phi_{\mu\nu}{}^{\rho\sigma}$ is a constant dual to the totally antisymmetric octonionic structure constant $\psi_{\mu\nu\rho}$ associated with $G_2$ [14][17]:

$$\phi_{4567} = \phi_{2374} = \phi_{1357} = \phi_{1276} = \phi_{2356} = \phi_{1245} = \phi_{1346} = +1 ~,\qquad (6.2a)$$

$$\phi_{\mu\nu\rho\sigma} \equiv +(1/3!)\epsilon^{\mu\nu\rho\sigma\tau\lambda\omega\psi}\psi_{\tau\omega\psi} ~,\qquad (6.2b)$$

$$\psi_{123} = \psi_{516} = \psi_{624} = \psi_{435} = \psi_{471} = \psi_{673} = \psi_{572} = +1 ~.\qquad (6.2c)$$

All other components, such as $\phi_{2357}$ are zero. So even though the conventional totally anti-symmetric $\epsilon$-tensor $\epsilon^{\mu\nu\rho\sigma}$ is absent in 7D due to the 4 indices fewer than 7, we still can define self-duality based on the reduced holonomy $G_2$ [15][16], using $\phi_{\mu\nu\rho\sigma}$.

Our objective now is to show that our system in 4D emerges out of a self-dual system in 7D, by a simple dimensional reduction. We consider the case of Euclidean 4D, because of the subtlety with the octonionic structure constant $\psi_{\mu\nu\rho}$ in the non-compact space-time $D = 4+3$ yielding $D = 2+2$ after a dimensional reduction.[9] Algebraically, the self-duality in 4D emerges out of the self-duality in 7D, because the holonomy $SO(4) \approx SU(2) \times SU(2)$ in 4D is a subgroup of the reduced holonomy $G_2$ in 7D [19].

For the purpose of a simple dimensional reduction 7D → 4D, we start with the self-duality conditions in 7D

$$\widehat{F}_{\hat\mu\hat\nu}{}^I \stackrel{*}{=} + \tfrac{1}{2} \widehat{\phi}_{\hat\mu\hat\nu}{}^{\hat\rho\hat\sigma} \widehat{F}_{\hat\rho\hat\sigma}{}^I ~,\quad \widehat{\mathcal{R}}_{\hat\mu\hat\nu}{}^I \stackrel{*}{=} + \tfrac{1}{2} \widehat{\phi}_{\hat\mu\hat\nu}{}^{\hat\rho\hat\sigma} \widehat{\mathcal{R}}_{\hat\rho\hat\sigma}{}^I ~.\qquad (6.3)$$

Needless to say, these self-dualities in 7D are also consistent with nilpotent fermionic symmetry, as has been mentioned after (2.7). From now on, we use the 'hat' symbols for the fields and indices in 7D, in order to distinguiush them from 4D fields and indices. To be more specific, we use the symbols $(\hat{x}^{\hat{\mu}}) = (x^\mu, y^\alpha)$ for the coordinates $x^\mu$ in 4D, and $y^\alpha$ in the extra three dimensions. The coordinate indices are now $(\hat{\mu}) = (4, 5, 6, 7; 1, 2, 3) = (\mu; \alpha)$.
[10]

---

[9] We are grateful to L. Borsten and M. Duff [18] for discussing this point.

The crucial requirements for our simple dimensional reduction are

$$\partial_\alpha \widehat{A}_{\hat{\mu}}{}^I \stackrel{*}{=} 0 \ , \quad \partial_\alpha \widehat{\psi}_{\hat{\mu}}{}^I \stackrel{*}{=} 0 \ , \quad \widehat{A}_\alpha{}^I \stackrel{*}{=} 0 \ , \quad \widehat{\psi}_\alpha{}^I \stackrel{*}{=} 0 \ , \tag{6.4a}$$

$$\widehat{F}_{\mu\alpha}{}^I \stackrel{*}{=} 0 \ , \quad \widehat{F}_{\alpha\beta}{}^I \stackrel{*}{=} 0 \ , \quad \widehat{\mathcal{R}}_{\mu\alpha}{}^I \stackrel{*}{=} 0 \ , \quad \widehat{\mathcal{R}}_{\alpha\beta}{}^I \stackrel{*}{=} 0 \ , \tag{6.4b}$$

$$\widehat{A}_\mu{}^I \stackrel{*}{=} A_\mu{}^I \ , \quad \widehat{\psi}_\mu{}^I \stackrel{*}{=} \psi_\mu{}^I \ , \quad \widehat{F}_{\mu\nu}{}^I \stackrel{*}{=} F_{\mu\nu}{}^I \ , \quad \widehat{\mathcal{R}}_{\mu\nu}{}^I \stackrel{*}{=} \mathcal{R}_{\mu\nu}{}^I \ , \tag{6.4c}$$

so that we are left up only with

$$F_{\mu\nu}{}^I \stackrel{*}{=} + \tfrac{1}{2} \epsilon_{\mu\nu}{}^{\rho\sigma} F_{\rho\sigma}{}^I \ , \quad \mathcal{R}_{\mu\hat{\nu}}{}^I \stackrel{*}{=} + \tfrac{1}{2} \epsilon_{\mu\nu}{}^{\rho\sigma} \mathcal{R}_{\rho\sigma}{}^I \ , \tag{6.5}$$

where $\epsilon_{\mu\nu}{}^{\rho\sigma} = \phi_{\mu\nu}{}^{\rho\sigma}$ is nothing but the epsilon tensor for 4D, because $\phi^{4567} = \epsilon^{4567} = +1$. In other words, we see that the self-duality conditions in (6.5) in 4D emerges out of self-duality conditions in (6.3) in 7D.

The only task left over is to confirm that our ansätze in (6.4) are actually consistent with the original self-duality conditions (6.3). This is rather easily done, as follows. First, for $(\hat{\mu}, \hat{\nu}) = (\mu, \alpha)$ in (6.3), the l.h.s. of $F$ and $\mathcal{R}$-equations are zero, while their r.h.s. also vanishes, because of the fact that $\widehat{\phi}_{\mu\nu\rho\alpha} = 0$ in (6.2). Second, for $(\hat{\mu}, \hat{\nu}) = (\alpha, \beta)$ in (6.3), there are only six independent equations

$$0 \stackrel{?}{=} \widehat{Y}_{12}{}^I = +\phi_{12}{}^{76} \widehat{Y}_{76}{}^I + \phi_{12}{}^{45} \widehat{Y}_{45}{}^I = Y_{76}{}^I + Y_{45}{}^I \ , \tag{6.6a}$$

$$0 \stackrel{?}{=} \widehat{Y}_{23}{}^I = +\phi_{23}{}^{74} \widehat{Y}_{74}{}^I + \phi_{23}{}^{56} \widehat{Y}_{56}{}^I = Y_{74}{}^I + Y_{56}{}^I \ , \tag{6.6b}$$

$$0 \stackrel{?}{=} \widehat{Y}_{31}{}^I = +\phi_{31}{}^{57} \widehat{Y}_{75}{}^I + \phi_{31}{}^{64} \widehat{Y}_{64}{}^I = Y_{75}{}^I + Y_{64}{}^I \ , \tag{6.6c}$$

where $Y$ is either $F$ or $\mathcal{R}$, in order to save space. The important fact is that these six equations are actually satisfied thanks to the six self-duality conditions (6.5) in 4D:

$$Y_{76}{}^I \stackrel{*}{=} + \epsilon_{76}{}^{45} Y_{45}{}^I = -Y_{45} \ , \tag{6.7a}$$

---

[10] The reason why we do not choose the simpler option, for example, $\mu = 1, 2, 3, 4$ and $\alpha = 4, 5, 6$ is due to the lack of the component $\phi_{1234} = +1$ in (6.2a), while $\phi_{4567}$ is non-zero for the four consecutive coordinates.

$$Y_{74}{}^I \stackrel{*}{=} + \epsilon_{74}{}^{56} Y_{56}{}^I = -Y_{56} ~~, \tag{6.7a}$$

$$Y_{75}{}^I \stackrel{*}{=} + \epsilon_{75}{}^{64} Y_{64}{}^I = -Y_{64} ~~. \tag{6.7a}$$

Notice that not only the self-duality of the Yang-Mills field strength $F_{\mu\nu}{}^I$ but also the self-duality of the vector-spinor field strength $\mathcal{R}_{\mu\nu}{}^I$ in 4D emerges out of the generalized self-duality in 7D. Note that these field strengths have non-trivial interactions due to the non-Abelian structure constants involved in these field strengths. We have to stress that such a system especially with a vector spinor has not been presented before, to our knowledge.

In principle, we can repeat similar confirmation for the dimensional reduction from the generalized self-duality 8D [15][16] into the self-duality in 4D, but we skip it in this paper.

## 7. Concluding Remarks

In this paper, we have given the system $(A_\mu{}^I, \psi_\mu{}^I, \chi^{IJ})$ with nilpotent fermionic symmetry in $D = 2+2$ with consistent interactions as in [7]. Our self-duality (3.1) is re-casted into (3.5), with hidden supersymmetry valid for supersymmetric integrable models in $D \leq 3$. Explicit examples are supersymmetric Kadomtsev-Petviashvili equations in $D = 2+1$ [11] and supersymmetric Korteweg-de Vries equations in $D = 1+1$ [13].

The emerging of hidden symmetries in lower dimensions is not new. For example, $N = 1$ supergravity in 11D yields the hidden symmetry $E_{7(+7)}/SU(8)$ after a dimensional reduction into 4D [8]. However, in the case of supersymmetry, it is usually reduced or preserved in dimensional reductions. Our system is a counter-example against such common observations, because the number of supercharges is increased in dimensional reductions.

We can generalize our result beyond $D = 2+2$ for the following reasons. First, our algebra (2.1) is valid in arbitrary space-time dimensions $D$. Second, our field strengths are defined by (2.2) in arbitrary $D$. Third, our transformations $\delta_N$ in (2.5) and $\delta_T$ in (2.6) are valid in arbitrary $D$. Fourth, our self-duality (2.8) is genaralized to higher-dimensions without upper limit for $D$:

$$F_{\mu\nu}{}^I \stackrel{*}{=} + \tfrac{1}{2} \phi_{\mu\nu}{}^{\rho\sigma} F_{\rho\sigma}{}^I ~~, \quad \mathcal{R}_{\mu\nu}{}^I \stackrel{*}{=} + \tfrac{1}{2} \phi_{\mu\nu}{}^{\rho\sigma} \mathcal{R}_{\rho\sigma}{}^I ~~, \tag{7.1}$$

with an appropriate constant $\phi_{\mu\nu}{}^{\rho\sigma}$, such as the octonionic structure constant [20] in 7D for the reduced holonomy $G_2 \subset SO(7)$, and in 8D for the reduced holonomy $SO(7) \subset SO(8)$ [21][22][23]. Needless to say, (7.1) has the nilpotent symmetry $N_{\underline{\alpha}}$, as the formulation in section 2 (originally from [7]) is valid in any space-time dimensions. If we can establish (7.1) and show that our self-duality (3.1) in 4D is obtained by a dimensional reduction, such a theory in certain $D$ is 'more fundamental' than our theory in 4D.

As a matter of fact, supersymmetric self-dual Yang-Mills theories in dimensions $D = 4$, $D = 5, 6, 7 \ (mod \ 4)$, $D = 8 \ (mod \ 4)$, $D = 9, 10, 11 \ (mod \ 4)$ have been discussed in [24]. As a matter of fact, the existence of the constant $\phi_{\mu\nu}{}^{\rho\sigma}$ in general space-time dimension $D$ is discussed based on stability group $H \subset SO(D)$ [24].

There are five important aspects in our results. First, a vector-spinor $\psi_\mu{}^I$ with nilpotent fermionic symmetry in 4D [7] is found to be important, because of its new application to self-dual Yang-Mills fields. Second, it is not necessary to use the multiplet $(1, 1/2)$ for self-dual supersymmetric Yang-Mills for our purpose. Third, our system of $(A_\mu{}^I, \psi_\mu{}^I, \chi^{IJ})$ is valid also in higher dimensions, supported by the explicit example in 7D. Fourth, we have shown that this self-dual system in 7D generates our original self-dual system in 4D by a simple dimensional reduction. Fifth, we have given the explicit examples of lower dimensional supersymmetric integrable systems in 3D and 2D emerging out of non-supersymmetric system in $D = 2 + 2$. To our knowledge, these examples have not been explicitly given in the past.

Especially, the last point is the most important aspect in this paper. According to common wisdom about dimensional reductions, any lower-dimensional supersymmetry is attributed to higher-dimensional supersymmetry. In particular, as mentioned above, the size of lower-dimensional supersymmetries is usually smaller than the corresponding supersymmetry in higher dimensions, because supersymmetries are supposed to be broken (or at most preserved) in dimensional reductions. A typical example is $0 \leq N \leq 8$ in 4D arising out of $N = 1$ supergravity in 11D, because $N = 1$ in 11D corresponds to $N = 8$ in 4D. Our system in this paper serves as a counter-example against such common understanding, because $N = 0$ in $D = 2 + 2$ yielded $N \geq 1$ in $D = 2 + 1$ or $D = 1 + 1$. Based on our results, it is natural to conjecture that similar systems exist in higher dimensions, even beyond 11D, because nilpotent fermionic symmetries has no upper limit for space-time dimensions.

We are grateful to L. Borsten, M. Duff and M. Günaydin for discussions. This work is supported in part by Department of Energy grant # DE-FG02-10ER41693.